\documentclass[aps,preprint,showkeys,nofootinbib]{revtex4-1}

\usepackage{amsmath}
\usepackage{amsfonts}
\usepackage{bm}
\usepackage{graphicx}
\usepackage{pgffor}

\newcommand{\Ref}[1]{Ref.~\cite{#1}}
\newcommand{\SEC}[2]{\section{\label{sec:#1}#2}}
\newcommand{\Sec}[1]{Sec.~\ref{sec:#1}}
\newcommand{\FIG}[2]{\caption{\label{fig:#1}#2}}
\newcommand{\Fig}[1]{Fig.~\ref{fig:#1}}
\newcommand{\EQ}[1]{\label{eq:#1}}
\newcommand{\Eq}[1]{Eq.~(\ref{eq:#1})}
\newcommand{\Eqs}[2]{Eqs.~({\foreach \next in #1 {\ref{eq:\next}, }}\ref{eq:#2})}

\newcommand{\pnt}[1]{\bm{#1}}
\newcommand{\avg}[1]{\langle#1\rangle}
\newcommand{\eqdef}{\overset{\mathrm{def}}{=}}

\newcommand{\DeltaS}{\Delta{}S}
\newcommand{\DeltaP}{\Delta{}P}

\begin{document}

\title{Asymmetry of steady state current fluctuations in nonequilibrium systems}
\date{\today}
\author{Roman Belousov}
\email{belousov.roman@gmail.com}
\author{E.G.D. Cohen}
\email{egdc@rockefeller.edu}
\affiliation{Rockefeller University}

\begin{abstract}
For systems in nonequilibrium steady states, a novel modulated Gaussian
probability distribution is derived to incorporate a new phenomenon of biased
current fluctuations, discovered by recent laboratory experiments and confirmed
by molecular dynamics simulations. Our results consistently extend
Onsager-Machlup fluctuation theory for systems in thermal equilibrium.
Connections with the principles of Statistical Mechanics due to Boltzmann and
Gibbs are discussed. At last, the modulated Gaussian distribution is of
potential interest for other statistical disciplines, which make use of the
Large Deviation theory.
\end{abstract}
\keywords{large deviation; probability distribution; asymmetry; current; nonequilibrium; fluctuation}
\maketitle

\SEC{intro}{Introduction}

In this letter the fluctuation theory of currents, originally proposed by
Onsager and Machlup for systems in thermal equilibrium
\Ref{OnsagerMachlup1953II}, is extended to the nonequilibrium Steady States
(SS). We suggest a novel Modulated Gaussian (MG) probability distribution, which
supersedes the normal distribution, used to describe the fluctuations of
thermodynamic variables in equilibrium. This new model features an asymmetry of
the current fluctuations, a property inherent to the nonequilibrium regime. MG
includes the Gaussian distribution as a special case so that the earlier theory
is consistent with ours. Furthermore, in \Sec{general} connections between our
formalism and the principles of statistical mechanics due to Boltzmann and Gibbs
are discussed.

Our research was motivated by the evidence of non-Gaussianity of current
fluctuations in a nonequilibrium SS, found in recent laboratory experiments
\Ref{UI} on a shear flow in a 2-Dimensional (2D) dusty plasma, similar to that
considered in \Ref{FGL2012}. The new probability distribution, which is based on
the Large Deviation (LD) approach (\Ref{Touchette2009}), allows to account for
this newly observed phenomenon. Our theoretical results are confirmed by
Molecular Dynamics (MD) simulations of a shear flow for a Debye-H\"{u}ckel (DH)
plasma.

The asymmetry of nonequilibrium SS current fluctuations, studied in this paper,
is due to a probability bias, namely due to the higher probability for the
current to deviate from its most likely value towards larger magnitudes, rather
than to smaller ones. Naturally, equilibrium systems are free of such a bias,
since their most likely magnitude of a current is zero \footnote{
	The temporal asymmetry predicted for trajectories of fluctuations onset and
	decay by the so called macroscopic fluctuation theory \Ref{MFT2015} is
	different from the bias of the time-independent nonequilibrium SS
	probabilities, studied here.
}.

Finally, we would like to remark, that the MG probability distribution, which is
derived in \Sec{mg} without any specific physical assumptions, relies solely on
the LD theory. Therefore, whenever the latter holds, the MG should also be
applicable, in principle, to random variables outside the field of Statistical
Mechanics.

\SEC{MD}{Computational model}

In our numerical experiments we studied an iso-thermal MD model of a 2D plasma
system. We will concentrate on computations performed with $N=100$ particles of
a unit mass $m$, interacting through the DH potential $\Phi_{DH}$, shifted and
truncated at a cutoff $r_c$:

\begin{eqnarray}
\Phi_{DH}(r) = \begin{cases}
	\epsilon \lambda \left[
    r^{-1} \exp(-\frac{r}{\lambda}) - r_c^{-1} \exp(-\frac{r_c}{\lambda})
  \right]\text{, if } r < r_c \\
	0\text{, if } r \ge r_c
\end{cases}\text{,}
\end{eqnarray}
where $r$ is the distance between two interacting particles, while the energy
constant $\epsilon$ and the Debye screening length $\lambda$ are chosen as the
basis of a reduced units system.

A constant shear rate and temperature of the plasma were maintained by the SLLOD
equations of motion coupled to the Nos\'{e}-Hoover (NH) thermostat in
$D=2$ dimensions (\Ref{EvansMorriss}):

\begin{eqnarray}\EQ{model}
\pnt{\dot{q}}_i &=& \frac{\pnt{p}_i}{m} + \gamma q_{iy} \pnt{X} \nonumber\\
\pnt{\dot{p}}_i &=& \pnt{F}_i(\pnt{q}_i)
	- \gamma p_{iy} \pnt{X} - \alpha_{NH} \pnt{p}_i \nonumber\\
\dot{\alpha}_{NH} &=& \theta^{-2} \left(
	\sum_{i=1}^{N}\frac{\pnt{p}_i^2}{D N k_B T m} - 1
\right)\text{.}
\end{eqnarray}

Here $\pnt{X}$ is a unit vector along the Cartesian x-coordinate axis, $k_B$ is
the Boltzmann constant, $\gamma = \frac{d u_x}{d y}$ is the applied shear rate,
{\it i.e.} the gradient of the streaming velocity $\pnt{u} = (u_x\,0)$; while
$\pnt{q}_i$, $\pnt{p}_i = m (\pnt{\dot{q}}_i - \pnt{u})$ and $\pnt{F}_i$ are,
respectively, the $i$-th particle position ($q_{iy}$ being its y-coordinate),
{\it peculiar} momentum and force on the $i$-th particle, resulting from the
interactions with all the other particles; finally, the NH thermostat is
characterized by its temperature $T$, the relaxation time $\theta$ and the
time-dependent coupling $\alpha_{NH}$. The flux of x-momentum along the
y-coordinate \footnote{
	{\it I.e.} the xy-component of the pressure tensor
}, generated by \Eq{model} in the primary simulation cell of the linear
dimension $L$, is

\begin{eqnarray}\EQ{pxy}
  P_{xy} = L^{-D} \sum_{i=1}^{N}\left(\frac{p_{ix} p_{iy}}{m} + F_{ix} y_i\right)
  \text{.}
\end{eqnarray}

In all our simulations $k_B T$ is set equal to $1\epsilon$, whereas $\theta$
was adopted as the unit of time. The equations of motion were integrated by the
classical 4th-order Runge-Kutta algorithm (\Ref{HairerWannerNorsett}), with the
time step $0.001\theta$ and the potential cutoff $r_c = 3\lambda$. The linear
size of primary cell $L$ was $8.86\lambda$.

\SEC{mg}{Modulated Gaussian distribution}
In this section we present a derivation of the MG distribution for the flux
\Eq{pxy}. We begin with a probability density function of the form:

\begin{eqnarray}\EQ{form}
  p(P_{xy}) = \exp\{\frac{S(P_{xy})}{k_B}\}\text{.}
\end{eqnarray}

The function $S(P_{xy})$, which depends implicitly on $N$, will be interpreted
in \Sec{general}. Considering each term of the summation operator in
\Eq{pxy} as a random variable, we assume, that their spatial average $P_{xy}$
obeys the LD theory. That is, there exists a rate function
\begin{eqnarray}\EQ{I}
	I(P_{xy}) = -\lim_{N\to\infty}\{S(P_{xy})/N k_B\}
\text{,}\end{eqnarray}
which has a global minimum
\begin{eqnarray}\EQ{ld}I(\tilde{P}_{xy})=0\end{eqnarray}
at the most likely value of $P_{xy} = \tilde{P}_{xy}$ ({\it cf.}
\Ref{Touchette2009}).

Expanding $S(P_{xy})$ in \Eq{form} about $\tilde{P}_{xy}$ in a power series up
to a prescribed finite order $n$, we obtain:

\begin{eqnarray}
  \EQ{series}
  \frac{S(P_{xy})}{k_B} &\approx& \frac{S(\tilde{P}_{xy})}{k_B}
	+ \sum_{i=1}^n \frac{S^{(i)}(\tilde{P}_{xy})}{i! k_B} (P_{xy}-\tilde{P}_{xy})^i
  \nonumber\\
  &\eqdef& -\frac{\tilde{S} + \DeltaS(\DeltaP_{xy})}{k_B}
\text{,}
\end{eqnarray}
which defines ($\eqdef$) a constant $\tilde{S}$ and a cost function
$\DeltaS(\DeltaP_{xy})$ of the deviation $\DeltaP_{xy}=P_{xy}-\tilde{P}_{xy}$,
which will be considered from a physical perspective in \Sec{general}.

From now on, the values of the derivatives $S^{(i)}(\tilde{P}_{xy})$ will be
denoted by $S_i$, for brevity. Since the cost function is zero for
$P_{xy} = \tilde{P}_{xy}$, \Eqs{I}{ld} suggest to pose:

\begin{eqnarray}\EQ{rate}
	\lim_{N\to\infty}\frac{\tilde{S}}{N k_B} &=& 0\nonumber\\
  I(P_{xy}) &\approx& -\lim_{N\to\infty}\frac{\DeltaS(\DeltaP_{xy})}{N k_B}
  \text{.}
\end{eqnarray}

As $p(\tilde{P}_{xy})$ is the global maximum of probability density, it follows
that the derivative $S_1$ vanishes. For a finite $N$ \Eq{form} then becomes:

\begin{eqnarray}\EQ{prb}
  p(P_{xy}) \approx \exp\{-\frac{\tilde{S}}{k_B} + \frac{\DeltaS(\DeltaP_{xy})}{k_B}\}
    = \exp\{-\frac{\tilde{S}}{k_B}
      + \sum_{i=2}^n\frac{S_i}{i! k_B}(\DeltaP_{xy})^i\}
  \text{,}
\end{eqnarray}

where the normalization of the total probability requires that

\begin{eqnarray*}
\exp\{\frac{\tilde{S}}{k_B}\} = \int_{-\infty}^{\infty} dP_{xy}
	\DeltaS(P_{xy} - \tilde{P}_{xy})
\text{.}\end{eqnarray*}

For $n=2$, \Eq{prb} turns into a Gaussian distribution. One may recognize that,
this case was treated by Onsager and Machlup in
\Ref{OnsagerMachlup1953I,OnsagerMachlup1953II} for the equilibrium state
($\tilde{P}_{xy}=0$) \footnote{
  Especially, current fluctuations about equilibrium are considered in
  \Ref{OnsagerMachlup1953II}.
}. Therefore to account for non-Gaussian phenomena, one must consider higher
order terms of the cost function, which can be arranged as follows:

\begin{eqnarray}\EQ{cost}
  \DeltaS(\DeltaP_{xy}) &=& \frac{S_2}{2 k_B} (\DeltaP_{xy})^2 +
    \sum_{i=3}^{n}\frac{S_i}{i! k_B} (\DeltaP_{xy})^{i}\nonumber\\
    &=& \frac{S_2}{2 k_B} (\DeltaP_{xy})^2 \left\{
      1 + 2 \sum_{i=3}^{n}\frac{S_i}{i! S_2} (\DeltaP_{xy})^{i-2}
    \right\}
  \text{,}
\end{eqnarray}
where the term in the curly braces
$\Sigma_n=1+2 \sum_{i=3}^{n}\frac{S_i}{i! S_2} \DeltaP_{xy}^{i-2}$ is a
modulating factor, to which the MG distribution owes its name.

The original Gaussian is recovered, when
$\Sigma_n \equiv 1$. For $p(P_{xy})$ to be integrable in \Eq{prb}, the order $n$
has to be restricted to even integers. Using the lowest order non-Gaussian
approximation, {\it i.e.} $n=4$, we replace the derivatives $S_i$ by three
parameters of scale $\Pi$, of asymmetry $A$, and of non-Gaussianity $B$, which
allow a more convenient interpretation, in \Eq{prb}:

\begin{eqnarray}\EQ{krn}
  p(P_{xy}) \propto \exp\{\DeltaS(\DeltaP_{xy})\} =
      \exp\{ \frac{S_2 (\DeltaP_{xy})^2}{2 k_B} \Sigma_4(\DeltaP_{xy})\}
    \nonumber\\
    = \exp\{ -\frac{(\DeltaP_{xy})^2}{2 \Pi^2} (
      1 - 2 \sqrt{2/3} A B \frac{\DeltaP_{xy}}{\Pi}
        + B^2 \frac{(\DeltaP_{xy})^2}{\Pi^2}
    )\}
  \text{.}
\end{eqnarray}

The dimensionless non-negative constant $B\ge0$ controls the level of
non-Gaussianity, thus $p(P_{xy})$ is exactly Gaussian for $B=0$ with the scale
parameter $\Pi$. $A$ determines the asymmetry of $p(P_{xy})$, which is skew to
the left (right), {\it i.e.} has a negative (positive) skewness, when $A<0$
($A>0$), respectively. The numerical factor $2\sqrt{2/3}$ in \Eq{krn}, as a
coefficient of the term with $A$, was chosen to make $\DeltaS(\DeltaP_{xy})$ a
non-concave function of $P_{xy}$, {\it i.e.}
$\frac{d^2 \DeltaS(\DeltaP_{xy})}{dP_{xy}^2} \ge 0$, when $-1 \le A \le 1$.
Violation of the non-concavity condition would admit special artifacts of the
probability distribution, {\it e.g.} a local maximum of its density. For
experimental or numerical observations, such flexibility may result in a data
over-fitting, and hence should be suppressed by using constrained optimization
techniques.

\SEC{hist}{Non-Gaussianity of current fluctuations}
\Fig{drb} illustrates probability density estimates \footnote{
  Everywhere we adopt the Freedman-Diaconis rule (\Ref{FreedmanDiaconis1981}) to
  calculate the bin width of traditional histograms and the band width of smooth
  histograms with the Gaussian kernel (\Ref{Silverman1986}).
} of the momentum current $P_{xy}$, at low and high shear rates, for our
computational experiments, including the fits of the Gaussian and modulated
Gaussian distributions (\Eq{krn}) \footnote{
	Because the normalizing constant of the MG can be evaluated only numerically,
	to obtain the corresponding Maximum Likelihood Estimator (MLE) of the MG
	model, we used the Wolfram Mathematica$^\text{\textregistered}$ implementation
	of the interior point method with the constraint of non-concavity, mentioned
	at the end of \Sec{mg}.
}. For the low value of $\gamma=0.2\theta^{-1}$ an excellent agreement is
observed between the shape of histograms and the parametric models
(indistinguishable on the graph for the lower shear rate). However, the Gaussian
model renders an unsatisfactory result in the case of a larger value
$\gamma=1.0\theta^{-1}$, where the superior performance of the MG distribution
is evident.

\begin{figure}[h]
\includegraphics[width=1\textwidth]{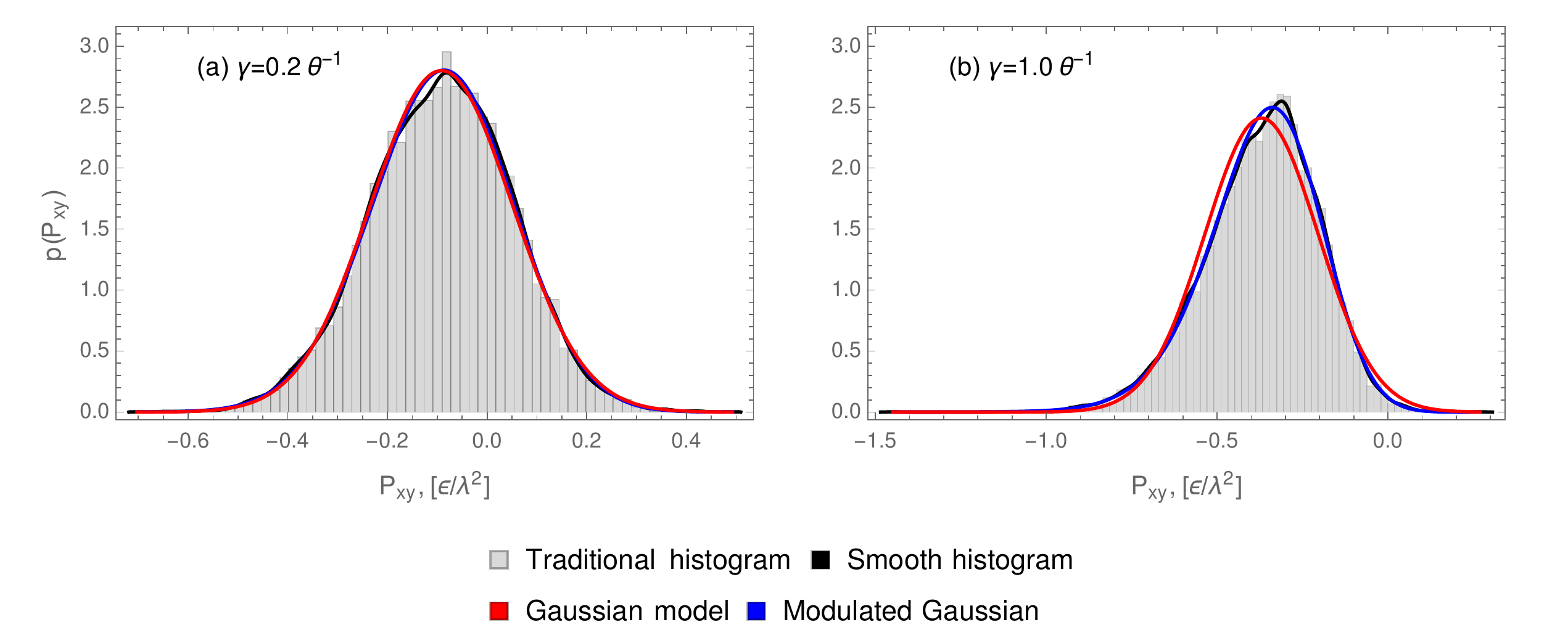}
\FIG{drb}{
  Probability density of $P_{xy}$: (a) for a low shear rate in the left panel;
	(b) for a high shear rate in the right panel. For the low shear rate the
	Gaussian and the MG are indistinguishable.
}
\end{figure}

Most obviously, the Gaussian model fails, when the fluctuations are asymmetric
about the mode of the probability distribution $\tilde{P}_{xy}$, as can be
quantified by the skewness statistics. Increasing the departure from equilibrium
augments the non-Gaussianity of $p(P_{xy})$, as confirmed by the plots of the
skewness and excess kurtosis, which are both zero for the normal distribution,
{\it vs.} the magnitude of the shear rate in \Fig{stats}. This occurs, because
in nonequilibrium the probability of a positive deviation $\DeltaP_{xy} > 0$
from the most likely value $\tilde{P}_{xy}$ becomes different from the
probability of the opposite deviation $-\DeltaP_{xy}$. \Fig{stats} also
demonstrates the efficiency of the modulated Gaussian to fit high-order
statistics.

\begin{figure}[h]
\includegraphics[width=1\textwidth]{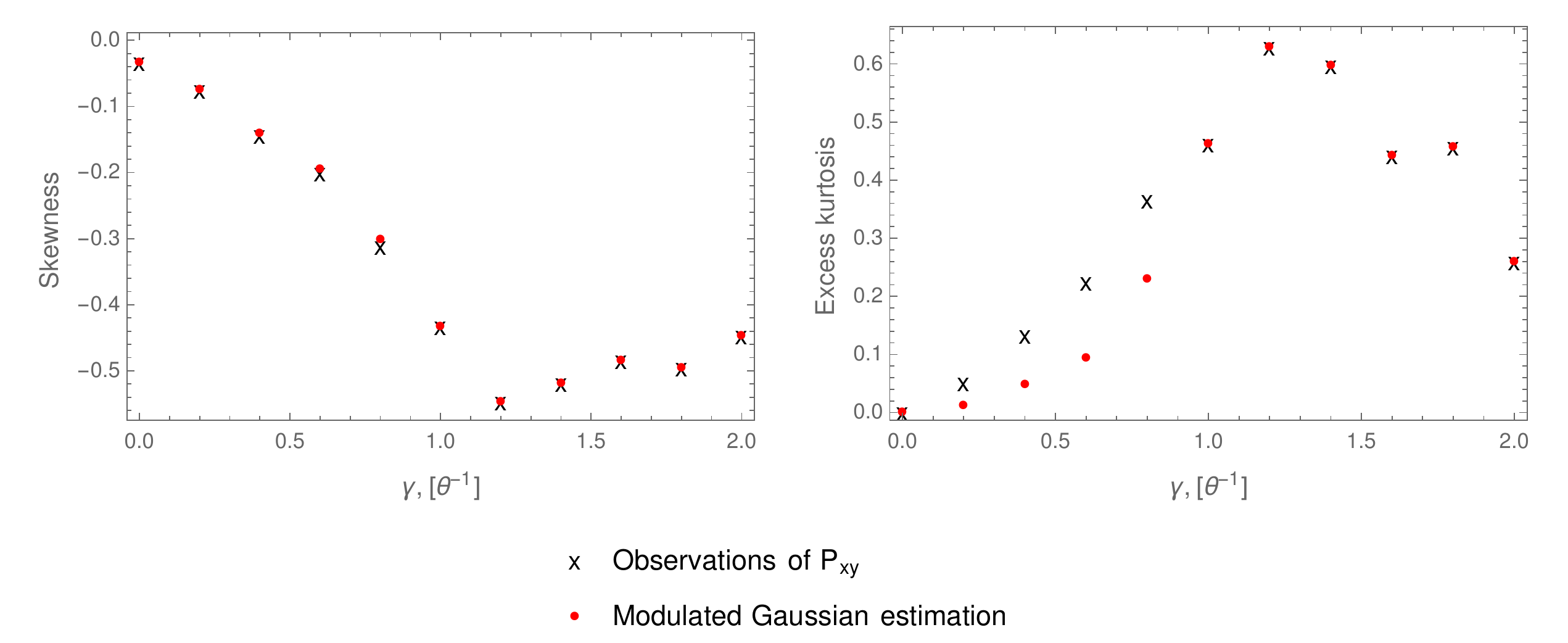}
\FIG{stats}{
  Statistics of skewness and excess kurtosis as a function of the shear rate for
	our observations of $P_{xy}$ and the corresponding fits by the MG.
}
\end{figure}

For a positive shear rate, which maintains a negative flux $P_{xy}$ ({\it cf.}
\Fig{drb}), we observe $p(\DeltaP_{xy}>0) < p(-\DeltaP_{xy})$.  The MG
distribution accounts for this bias of probabilities and fits very well the 3rd
and 4th order statistics (skewness and kurtosis, respectively). We conjecture,
that the asymmetry of the distribution $p(P_{xy})$ is a characteristic property
of the current fluctuations about a nonequilibrium SS. It is small, though, in
the near-equilibrium SS regime and vanishes only in the limit $\gamma\to 0$.

\SEC{general}{Connection with classical Statistical Mechanics}

The results obtained in \Sec{mg}, can be readily connected with the Boltzmann
and Gibbs principles of statistical mechanics. According to the Boltzmann
principle, given a measure of system microstates $w(P_{xy})$, for a given value
of $P_{xy}$, and the total measure of the accessible microstates
$W = \int_{-\infty}^{\infty} w(P_{xy}) dP_{xy}$ under the specified macroscopic
constraints of temperature, volume, shear rate {\it etc.}, the SS probability
density $p(P_{xy})$ and the Boltzmann entropy $S_B(P_{xy})$ are given by:

\begin{eqnarray}\EQ{sb}
p(P_{xy}) &=& \frac{w(P_{xy})}{W}\nonumber\\
S_B(P_{xy}) &=& k_B \ln w(P_{xy})
\text{,}\end{eqnarray}
respectively.

Using the notion of total entropy $S_{tot} = k_B \ln W$ after \Ref{Attard}, we
deduce from \Eq{sb} that:

\begin{eqnarray}\EQ{log}
k_B \ln p(P_{xy}) &=& k_B \ln w(P_{xy}) - k_B \ln W
	= S_B(P_{xy}) - S_B(\tilde{P}_{xy}) + S_B(\tilde{P}_{xy}) - S_{tot}
	\nonumber\\
	&=& \DeltaS_B(P_{xy}) + S_B(\tilde{P}_{xy}) - S_{tot}
\text{,}\end{eqnarray}
where in the second equality we added and subtracted $S_B(\tilde{P}_{xy})$ to
introduce the Boltzmann entropy difference
$\DeltaS_B(P_{xy}) = S_B(P_{xy}) - S_B(\tilde{P}_{xy})$.

Comparing \Eq{prb} with \Eq{log}, one sees that

\begin{eqnarray}\EQ{contact}
	\tilde{S} &=& S_{tot} - S_B(\tilde{P}_{xy})
	\nonumber\\
	\DeltaS(P_{xy}-\tilde{P}_{xy}) &=& \DeltaS_B(P_{xy})
\text{,}\end{eqnarray}
because $\DeltaS_B(P_{xy})$ and $\DeltaS(P_{xy}-\tilde{P}_{xy})$ are both zero
at $P_{xy} = \tilde{P}_{xy}$ by definition.

Hence Eq.~(\ref{eq:contact}) provides the interpretation of the cost function
$\DeltaS(\DeltaP_{xy})$ and the constant $\tilde{S}$, introduced in \Sec{mg}, in
terms of Boltzmann entropy and the total entropy. The most likely value of
$P_{xy}=\tilde{P}_{xy}$ maximizes the Boltzmann entropy ({\it cf.} \Eq{sb}).
Consistently, $\DeltaS(\DeltaP_{xy})\le 0$ is the (Boltzmann) entropy cost of a
fluctuation $P_{xy} = \tilde{P}_{xy} + \DeltaP_{xy}$ \footnote{
  One can also define the (free) energy cost of a fluctuation
	$\Delta{}F=\DeltaS(P_{xy}) T$, {\it cf.} \Ref{LandauLifshitz}.
}. The constant $\tilde{S}$ is the remaining total entropy, after subtracting
the entropy cost of the most likely macrostate.

Finally, the Gibbs entropy, given by a functional $S_G[p]$, for the distribution
\Eq{prb} is

\begin{eqnarray}\EQ{sg}
	S_G[p(\cdot)] &= -k_B \int_{-\infty}^{\infty} dP_{xy} p(P_{xy}) \ln p(P_{xy}) =
		-k_B \avg{\ln p(P_{xy})} \nonumber\\
		&= \avg{\tilde{S} - \DeltaS(P_{xy})} = \tilde{S} - \avg{\DeltaS(P_{xy})}
		\nonumber\\
		&= S_{tot} - S_B(\tilde{P}_{xy}) - \avg{\DeltaS(P_{xy})}
\text{,}\end{eqnarray}

which extracts $\tilde{S}$, up to a constant term $\avg{\DeltaS(P_{xy})}$, from
the SS probability density $p(P_{xy})$.

\SEC{fin}{Conclusion}

In \Sec{mg} we used the 4th order modulating factor ($\Sigma_4$) to derive
our modulated Gaussian probability distribution. In principle, the same
procedure can be applied to obtain the next order approximation, using
$\Sigma_6$. However, the formalism becomes then much more complicated. The
performance of the forth order factor was demonstrated in \Sec{hist} and proved
to be sufficient to describe the asymmetry of fluctuations and the decay of
their probability distribution tails. Although we presented our data only for
the simulations with $N=100$ particles, we checked that the modulated Gaussian
performs equally well for smaller and larger values $N$ as well.

An important consequence of the asymmetric fluctuations about the SS is the
discrepancy between the current average and its most likely value $\avg{P_{xy}}
\ne \tilde{P}_{xy}$, which are equal in equilibrium. Since we do not provide the
dynamics of fluctuations, like the Langevin equation suggested by Onsager and
Machlup for equilibrium, it is an open question, how this probability bias
emerges. There are various ways to modify the Langevin equation to produce
asymmetric time-independent distribution, {\it e.g.} non-Gaussian noise or
non-linear deterministic term. Physically, this could be a consequence of the
arrow of time in a nonequilibrium stystem, which admits a bias. Yet a dynamical
theory remains to be found.

\bibliographystyle{plain}
\bibliography{References}

\end{document}